\documentclass{mn2e}
\usepackage{graphicx}
\newcommand{\Fs}{\,^*\! F}

\newcommand{\bB}{\bmath{B}}

\newcommand{\text}[1]{\quad\mbox{#1}\quad}

\begin{document}
\title{Magnetized Tori around Kerr Black Holes: Analytic Solutions with 
a Toroidal Magnetic Field} 
\author[S.S.Komissarov]{S.S.Komissarov\\
Department of Applied Mathematics, the University of Leeds,
Leeds, LS2 9GT, UK.\\
e-mail: serguei@maths.leeds.ac.uk}
\date{Received/Accepted}
\maketitle
                                                                                                
\begin{abstract}

The dynamics of accretion discs around galactic and extragalactic black holes 
may be influenced by their magnetic field. In this paper we generalise 
the fully relativistic theory of stationary axisymmetric tori in Kerr metric 
of Abramowicz et al.\shortcite{Abr78} by including strong toroidal magnetic field
and construct analytic solutions for barotropic tori with constant angular 
momentum. This development is particularly important for the general relativistic 
computational magnetohydrodynamics that suffers from the lack of exact analytic 
solutions that are needed to test computer codes.  
\end{abstract}

\begin{keywords}
black hole physics -- accretion discs -- MHD -- methods:analytical -- 
methods:numerical
\end{keywords}
                                                                                                
\section{Introduction}

Accretion discs of black holes have been a subject of intensive 
observational and theoretical studies for decades. In most of the studies 
the discs are treated as purely fluid flows \cite{ShS,FiM,Abr78,Ree}. 
That is their magnetic fields are considered as dynamically weak and not 
important for the force balance determining the disc structure. On the other,
it has been long recognised that MHD-turbulence may hold the key to the 
nature of the disc viscosity that enables the inflow of matter into 
the black hole \cite{ShS}. This expectation was confirmed with the discovery 
of the magneto-rotational instability \cite{BaH} that has become now one of 
the major topics in theoretical and numerical studies of accretion discs. 
In the simulations of radiatively inefficient discs they remain 
relatively weakly magnetized with the magnetization parameter 
$\beta =(\mbox{gas pressure})/(\mbox{magnetic pressure}) \simeq 10-100$ 
within the main body of the disc and rising up to $\beta \simeq 1$ only near 
its inner edge and in the plunging region \cite{HKDH,MG}. However, cooling 
accretion flows may well evolve towards 
the state with $\beta<1$ where the vertical balance against the 
gravitational force of the central object is achieved by means of 
magnetic pressure alone \cite{MNM,PBB}.        

On the other hand, there is a general consensus that dynamically 
strong, ordered, poloidal magnetic field is required both for  acceleration and 
collimation of the relativistic jets that are often produced by the
astrophysical black hole-accretion disc systems. Such magnetic field 
is a key ingredient both in the models where the jet is powered 
by the black hole \cite{BZ} and in the models where it is powered by the 
accretion disc \cite{BR,BP}.
The origin of this field is not very clear. The most popular idea is that  
it is carried into central parts of the disc-hole system by the accreting 
flow itself and that the net magnetic flux gradually builds up there during 
the long-term evolution of the system \cite{BR,TM}.  This is more or 
less what is observed in recent numerical simulations of radiatively 
inefficient discs with initially weak poloidal field, \cite{HKDH,MG}. 
However, the strength of magnetic field that can be reached in this way 
is still unclear, e.g. \cite{LOP,SU,Meier,McK}. 
The most extreme case, where the disc pressure is 
dominated by ordered poloidal magnetic field is argued by 
Meier \shortcite{Meier} as a model for the low/hard state of X-ray binaries.    
Although much weaker magnetic field is required to explain the observed 
power of quasar jets within the Blandford-Znajek theory,  the 
pressure of poloidal magnetic field still has to be of the same order as the 
radiative pressure at the inner edge of the radiatively supported accretion discs 
\cite{BBR}. Bogovalov and Kel'ner \shortcite{BK} constructed a model 
of accretion disc where its angular momentum is carried away by a magnetized 
wind alone and the inflow of matter and magnetic field is driven entirely by 
the magnetic torque applied to the disc by the wind. These and other results 
show that the structure of accretion discs with dynamically strong 
ordered magnetic is a matter of significant astrophysical interest.  

General steady-state axisymmetric magnetized dicks are described by a 
very complex equation of the Grad-Shafranov type which does not allow 
general analytic solutions \cite{Lov}. However, the equilibrium models with 
pure azimuthal  (toroidal) field are much simpler and can be constructed using 
the approach developed for unmagnetised discs \cite{Abr78,Koz}. In fact, the 
predominant motion in accretion discs is a differential rotation and one 
would expect the azimuthal magnetic field to dominate in the interior of such 
discs.    
 
Okada et al.\shortcite{Oka} constructed a particular exact solution 
for the problem of equilibrium torus with purely azimuthal magnetic 
field.  In their approach the Paczynski \& Wiita \shortcite{PW} potential 
is utilised to introduce the black hole gravity and the flow dynamics is described
by the non-relativistic equations. Although this approach seems to work fine 
in the case of non-rotating black holes, it is no longer satisfactory in  
the astrophysically important case of rapidly rotating black hole where 
full relativistic treatment is required. In this paper 
we develop the general relativistic theory of magnetized tori around 
rotating black holes. Our analysis closely follows the one of 
Kozlowski et al.\shortcite{Koz} and Abramowicz et al.\shortcite{Abr78}, 
including the notation.

Even if tori with dynamically strong magnetic fields do not exist in nature 
the exact solutions presented here can still be useful for numerical 
general relativistic magnetohydrodynamics, GRMHD, which has attracted a 
lot of interest recently \cite{KSK,Kom01,DvH,GMT,Kom,Due,Ant}.    
One of the problems with general relativistic computational 
magnetohydrodynamics is the lack of exact analytic and semi-analytic solutions 
that could be used to verify computer codes. The current list of such solutions 
includes 
1) the spherical accretion onto a nonrotating black hole with 
monopole magnetic field where the magnetic field is dynamically passive 
\cite{KSK}, 
2) the perturbative force-free solution of Blandford-Znajek \shortcite{BZ}, 
where both the pressure and the inertia of matter is neglected and the 
black hole rotates slowly, and 
3) the so-called Gammie flow, that is a one-dimensional solution 
of the Weber-Davis type that applies only to the equatorial plane of a 
black hole \cite{Gam}. 
One may also try to utilize the axisymmetric semi-analytic self-similar 
solutions constructed recently by Meliani et al.\shortcite{Mel} but those 
apply only along the symmetry axis.  
Obviously, the magnetized torus solution is very welcome an addition to this 
scarce list. In fact, it is the only solution that is not only 
a) fully multi-dimensional and b) involves dynamically important magnetic field, 
but c) also applies in the case of a rapidly rotating black hole.  
In fact, we have already used this solution to test the 2D GRMHD codes 
described in Komissarov \shortcite{Kom01,Kom}.

Throughout the paper we use the relativistic units where $c=G=M=1$ 
and $(-+++)$ signature for the space-time geometry. Following Anile \shortcite{Ani}
the magnetic field is rescaled in such a way that the factor $4\pi$ disappears
from the equations of relativistic MHD. 
\section{Basic equations}

The covariant equations of ideal relativistic MHD are 
                                                                                                
\begin{equation}
  \nabla_\alpha T^{\alpha \beta} = 0,
\label{em-eq}
\end{equation}
                                                                                                
\begin{equation}
  \nabla_\alpha \Fs^{\alpha \beta} = 0,
\label{Faraday-eq}
\end{equation}
                                                                                                
\begin{equation}
  \nabla_\alpha \rho u^\alpha = 0
\label{continuity-eq}
\end{equation}
where
                                                                                                
\begin{equation}
  T^{\alpha \beta} = (w+b^2) u^\alpha u^\beta +
                      (p+{1 \over 2} b^2) g^{\alpha \beta}
                      - b^\alpha b^\beta
\label{sem-tensor}
\end{equation}
                                                                                                
\noindent
is the energy-momentum tensor,   $w$, $p$ and  $u^\alpha$ are
the fluid enthalpy, pressure and  4-velocity of plasma respectively, 
and $g_{\alpha \beta}$ is the metric tensor \cite{Dix,Ani}.
                                                                                                
\begin{equation}
 \Fs^{\alpha \beta}  = b^\alpha u^\beta - b^\beta u^\alpha
\label{Faraday}
\end{equation}
                                                                                                
\noindent
is the Faraday tensor and  $b^\alpha$ is the 4-vector of magnetic field. 
In the fluid frame $b^\alpha = (0,\bB)$, where $\bB$ is the
the usual 3-vector of magnetic field as measured in this frame, and thus 

\begin{equation}
 u^\alpha b_\alpha =0.
\label{constraint}
\end{equation}

\noindent  
In the following we assume that  
\begin{enumerate}
\item
the space-time is described by the Kerr 
metric and that $\{t,\phi,r,\theta\}$ are either Kerr-Schild or 
Boyer-Lindquist coordinates, so that 
                                                                                                      
\begin{equation}
    g_{\mu\nu,t}=g_{\mu\nu,\phi} =0;   
\label{Kerr-m}
\end{equation}

\item
the flow is both stationary and axisymmetric, so that  

\begin{equation}
   f_{,t}=f_{,\phi} =0
\label{axi-sta}
\end{equation}
for any physical parameter $f$;

\item
the flow is a pure rotation around the black hole, that is 

\begin{equation}
    u^r=u^\theta=0;
\label{azim-u}
\end{equation}

\item 
the magnetic field is purely azimuthal, that is 
\begin{equation}
    b^r=b^\theta=0.
\label{azim-b}
\end{equation}
\end{enumerate}

In terms of partial derivatives the continuity equation reads 
$$
     (\sqrt{-g}\rho u^\nu)_{,\nu} = 0,
$$
where $g$ is the determinant of the metric tensor. Given the symmetry conditions 
(\ref{Kerr-m}) and (\ref{axi-sta}) this equation reduces to 
$$
     (\sqrt{-g}\rho u^i)_{,i} = 0,
$$
where here and throughout the whole paper $i=r,\theta$. Finally, the condition 
(\ref{azim-u}) tells us that this equation is always satisfied.  

Since the Faraday tensor is antisymmetric one may reduce the Faraday equation 
to 
$$
   (\sqrt{-g} \Fs^{\mu\nu})_{,\nu} = 0.
$$
It is easy to see that this equation is also automatically satisfied given 
the conditions (\ref{Kerr-m}-\ref{azim-b}).

Thus, the only non-trivial results follow from the energy-momentum equation (\ref{em-eq}). 
Contracting this equation with the projection tensor 
$h^{\alpha}_{\ \beta} = \delta^{\alpha}_{\ \beta} +u^\alpha u_\beta $
we obtain                                                                                                       
\begin{equation}
    (w+b^2)u_\nu u^\nu_{\ ,i} +
    (p+b^2)_{,i} - b_\nu b^\nu_{\ ,i} =0.
\label{eq-1}
\end{equation}

\noindent
In terms of the angular velocity, 
\begin{equation}
\Omega = u^\phi/u^t,
\label{eq-2}
\end{equation}
and the specific angular momentum 
\begin{equation}
l=-u_\phi/u_t.
\label{eq-3}
\end{equation}
(both $u_t$ and $u_\phi$ are constants of geodesic motion) this equation 
reads

\begin{equation}
    (\ln |u_t|)_{,i} -\frac{\Omega}{1-l\Omega} l_{,i}  
      + \frac{p_{,i}}{w} +
    \frac{({\cal L}b^2)_{,i}} {2{\cal L}w} =0,  
\label{eq-4}
\end{equation}
where 

\begin{equation}
{\cal L}(r,\theta) = g_{t\phi}g_{t\phi}-g_{tt}g_{\phi\phi}.
\label{eq-5}
\end{equation}
When $b^2 \to 0$ equation (\ref{eq-4}) reduces to equation 7 in 
Abramowicz et al.\shortcite{Abr78} 
which describes equilibrium non-magnetic tori. Thus, the Lorentz force vanishes 
and we have a force-free magnetic torus provided  
$$
  {\cal L} b^2 = \mbox{const}.
$$    
Far away from the black hole $g_{t\phi} \to 0$, $g_{\phi\phi} \to r^2\sin^2\theta$, 
$g_{tt}\to -1$ and this equation reduces to 
the familiar 
$$
   B^{\hat{\phi}} = \frac{\mbox{const}}{r\sin\theta}. 
$$

\section{Integrability conditions}

For a barotropic equation of state, where 
\begin{equation}
w=w(p), 
\label{eq-5a}
\end{equation}
\noindent
equation (\ref{eq-4}) leads to  
\begin{equation}
 d\left(\ln |u_t| +\int\limits_0^p\frac{dp}{w}\right) = 
        \frac{\Omega}{1-l\Omega} dl  - 
         \frac{d({\cal L}b^2)}{2{\cal L}w}
\label{eq-6}
\end{equation} 
In the case of a non-magnetic torus this equations implies that 
\begin{equation}
\Omega=\Omega(l)
\label{eq-6a}
\end{equation}
and thus the surfaces of equal $\Omega$, $l$, $p$, 
and $\rho$ coincide \cite{Abr71,Abr78}.  Obviously, this does not have 
to be the case for magnetized tori. If, however, we still assume 
that $\Omega=\Omega(l)$ then eq.(\ref{eq-6}) can be written as 
      
\begin{equation}
 d\left(\ln |u_t| +\int\limits_0^p\frac{dp}{w} - 
        \int\limits_0^l\frac{\Omega dl}{1-l\Omega} \right) =
         -\frac{d({\cal L}b^2) }{2{\cal L}w} 
\label{eq-7}
\end{equation} 
that implies that the expression of the right-hand side is a total 
differential. Hence, 
\begin{equation}
 \tilde{w}= \tilde{w}(\tilde{p}_m), 
\label{eq-7a}
\end{equation}
where $\tilde{w}={\cal L} w$ and  $\tilde{p}_m = {\cal L} p_m$, where 
$p_m=b^2/2 $ is the magnetic pressure, and eq.(\ref{eq-7}) integrates to give 
\begin{equation}
   \ln |u_t| +\int\limits_0^p\frac{dp}{w} -
   \int\limits_0^l\frac{\Omega dl}{1-l\Omega} +
   \int\limits_0^{\tilde{p}_m} \frac {d\tilde{p}_m}{\tilde{w}} = 
   \mbox{const}
\label{eq-8}
\end{equation}

Assuming that on the surface of the disc, and hence on its inner 
edge,  
\begin{equation}
  p=p_m=0, \quad u_t=u_{t_{in}}, \quad l=l_{in}
\label{eq-9}
\end{equation}
one finds the constant of integration as 
\begin{equation}
   \mbox{const} = \ln|u_{t_{in}}| -
     \int\limits_0^{l_{in}} \frac{\Omega dl}{1-l\Omega} . 
\label{eq-10}
\end{equation}
Following, Abramowicz et al.\shortcite{Abr78} we introduce the 
total potential, $W$, via 
\begin{equation}
  W = \ln|u_t| + \int\limits_l^{l_\infty} \frac{\Omega dl}{1-l\Omega},
\label{eq-11}
\end{equation}
where $l_\infty$ is the angular momentum at infinity. 
Provided  that $l_\infty$ is finite we have  $u_{t_{\infty}} = -1$ and 
$W_\infty = 0$. Using the total potential we can rewrite eq.(\ref{eq-8}) 
as 
\begin{equation}
   W-W_{in} +  
   \int\limits_0^p\frac{dp}{w} + 
   \int\limits_0^{\tilde{p}_m} \frac {d\tilde{p}_m}{\tilde{w}} = 0. 
\label{eq-11a}
\end{equation}
With exception for the last term this equation is the same as 
eq.9 in Abramowicz et al.\shortcite{Abr78} and eq.24 in 
Kozlowski et al.\shortcite{Koz}.   


\section{Barotropic tori with constant angular momentum}

\subsection{Theory}

Here we adopt particular relationships $w=w(p)$, $\Omega=\Omega(l)$, 
and $\tilde{w}=\tilde{w}(\tilde{p}_m)$ that allow to express the integrals 
of equation (\ref{eq-11a}) in terms of elementary functions. Namely, we 
assume that 
\begin{equation}
   l=l_0,
\label{eq-12}
\end{equation}  
\begin{equation}
   p=K w^\kappa
\label{eq-13}
\end{equation}  
\begin{equation}
   \tilde{p}_m = K_m \tilde{w}^\eta.
\label{eq-14}
\end{equation}  
The last equation can also be written as
\begin{equation}
   p_m = K_m {\cal L}^{\eta-1} w^\eta.
\label{eq-14a}
\end{equation}  

\noindent
Then eq.(\ref{eq-11a}) reduces to 
\begin{equation}
   W-W_{in} + \frac{\kappa}{\kappa-1} \frac{p}{w} + 
              \frac{\eta}{\eta-1} \frac{p_m}{w} = 0,  
\label{eq-15}
\end{equation}
where 
\begin{equation}
  W = \ln|u_t|. 
\label{eq-16}
\end{equation}

The obvious parameters of the model are $\kappa$, $\eta$, $l_0$, and 
$W_{in}$. Two more parameters are needed. We chose these to be the 
enthalpy, $w_c$, and the magnetization parameter $\beta_c = (p/p_m)_c$ at the 
disc centre, $r=r_c$. Following Abramowicz et al.\shortcite{Abr78}
we define the disc centre as one of the two points in the equatorial plane  
where $l_0$ equals to the Keplerian angular momentum

\begin{equation}
  l_k=\frac{\pm(r^2 \mp 2ar^{1/2}+a^2)}{r^{3/2}-2r^{1/2} \pm a} ,
\label{eq-17}
\end{equation}
where the upper sign is used if $l_0>0$ and the lower sign otherwise
\cite{Bar}.  The second point is to the disc ``cusp'', $r_{cusp}<r_c$  
\cite{Abr78}.  There exist a number of obvious constraints on the values 
of these parameters. 

In order to avoid divergence of the second and/or the fourth terms in (\ref{eq-15}) 
at the disc surface one should have 
$$
   \kappa ,\beta > 1.
$$   
Under this condition the disc surface is fully 
determined by the choice of $W_{in}$ and does not depend on the disc 
magnetization.  This property has already been noticed in Okada et 
al.\shortcite{Oka}. 

Only if 
$$
|l_0| > |l_{ms}|,
$$ 
where $l_{ms}$ is the radius of the marginally stable 
Keplerian orbit \cite{Bar}, the disc is detached from the event horizon
(Abramowicz et al. \shortcite{Abr78}. Solutions attached to the event horizon 
are improper because they diverge at the event horizon and cannot be 
continued through it.) 
The value of $l_0$ determines the total potential which can be written as
\begin{equation}
  W(r,\theta)=\frac{1}{2}\ln\left|\frac{\cal L}{\cal A} \right|, 
\label{eq-18}
\end{equation}
where 
$$  
    {\cal L}  = g_{t\phi}g_{t\phi} -g_{tt}g_{\phi\phi},
$$  
and
$$  
    {\cal A}  = g_{\phi\phi}+2l_0g_{t\phi}+ l_0^2 g_{tt}
$$  
(see eq.\ref{eq-a9} ). 
    
If $|l_0| \ge |l_{mb}|$ then the disc has finite outer radius only for 
$$
W_{in}<0.
$$ 

If $|l_{ms}|<|l_0|<|l_{mb}|$ where $l_{mb}$ is the angular momentum of 
the marginally bound Keplerian orbit \cite{Bar}, then the disc remains detached 
from the black hole only if 
$$W_{in} \le W_{cusp},$$ 
where $W_{cusp}$ is the value of the total potential at the cusp \cite{Abr78}.  
 
From eq.(\ref{eq-15}) one finds the gas pressure 
\begin{equation}
   p_c = w_c (W_{in}-W_c) 
  \left(\frac{\kappa}{\kappa-1} + 
        \frac{\eta}{\eta-1} \frac{1}{\beta_c} 
   \right)^{-1} ,  
\label{eq-19}
\end{equation}
and then the magnetic pressure  

\begin{equation}
   p_{m_c} =  p_c/ \beta_c
\label{eq-20}
\end{equation}
at the disc centre. Using these one finds the constants $K$ and $K_m$ of the 
barotropics (\ref{eq-13},\ref{eq-14a}). 

Now one can compute the solution at any location inside the disc. 
Given the coordinates  $(r,\theta)$  one computes the potential $W$ 
using equation (\ref{eq-18}). If $W<W_{in}$ then this point is inside 
the disc. The next step is to find the enthalpy, $w$, as a solution of 
eq.(\ref{eq-15}). Once $w$ is found one can find $p$ and $p_m$ from 
eqs.(\ref{eq-13}) and (\ref{eq-14a}).  The 4-velocity vector, 
$u^\nu=(u^t,u^\phi,0,0)$, is given by 
$$
    u^t = -\frac{1}{u_t(1-l_0\Omega)}, \quad u^\phi = \Omega u^t, 
$$
where $\Omega$ can be found via
$$
   \Omega= -\frac{g_{t\phi} + g_{tt} l_0}
            {g_{\phi\phi} + g_{t\phi} l_0 }
$$ 
(see Appendix.) The 4-vector of magnetic field, $b^\nu=(b^t,b^\phi,0,0)$, 
is given by 
$$
 b^\phi = \pm\sqrt{2p_m/{\cal A}}, \quad b^t = l_0 b^\phi  
$$                                                                  
(see eq.\ref{eq-a13a}).
                                                                        

\begin{figure*}
\fbox{\includegraphics[width=55mm]{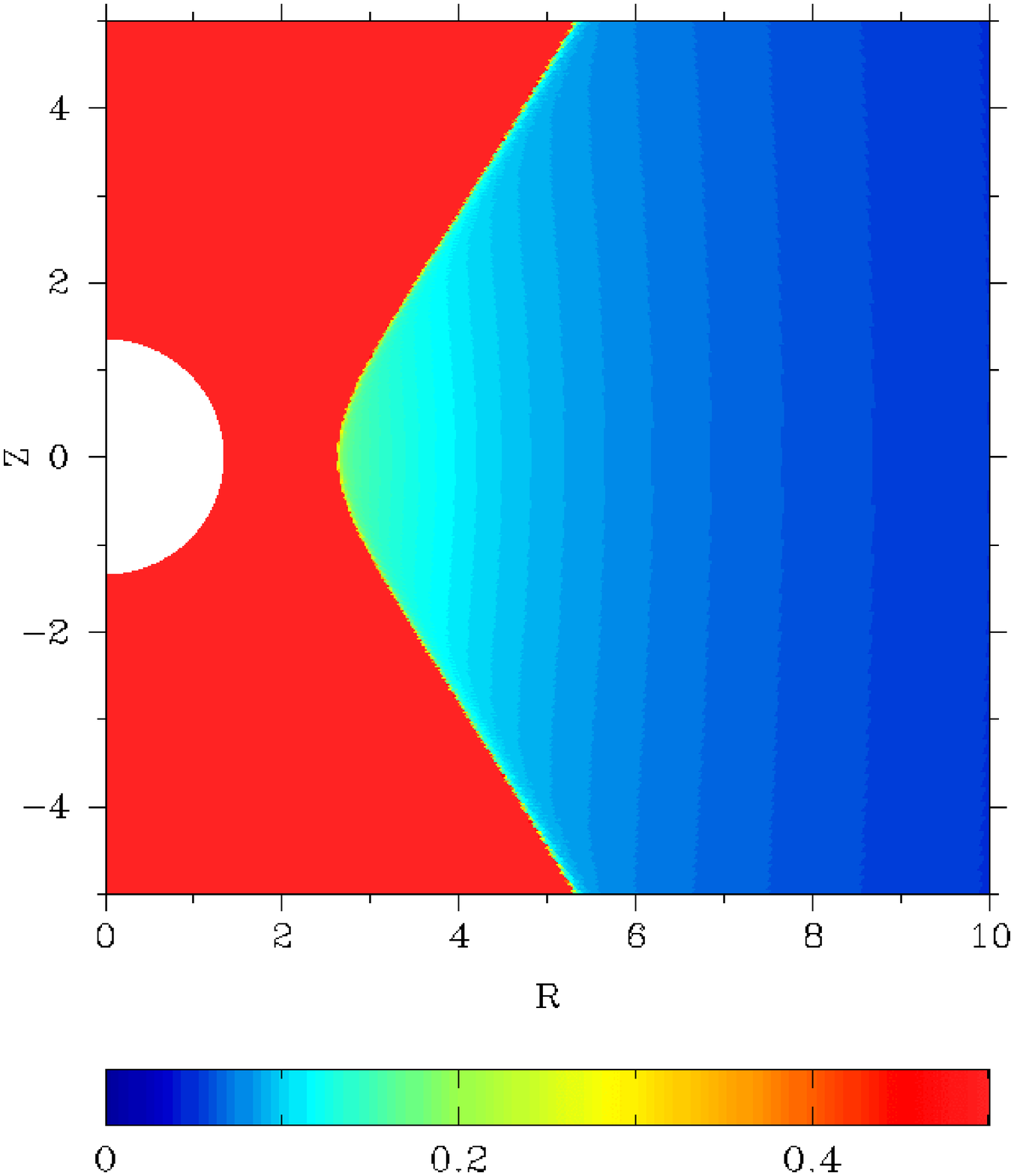}}
\fbox{\includegraphics[width=55mm]{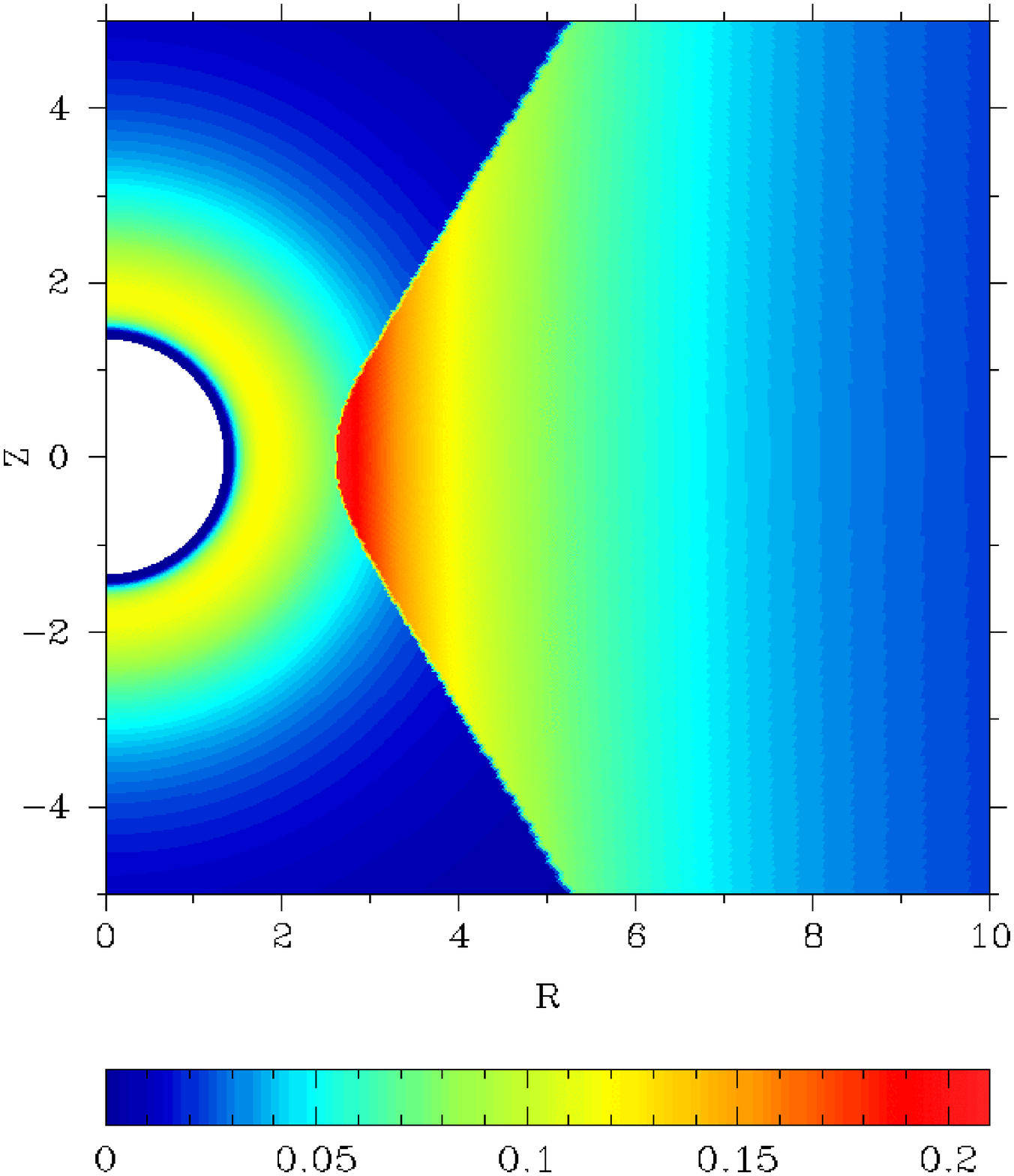}}
\fbox{\includegraphics[width=55mm]{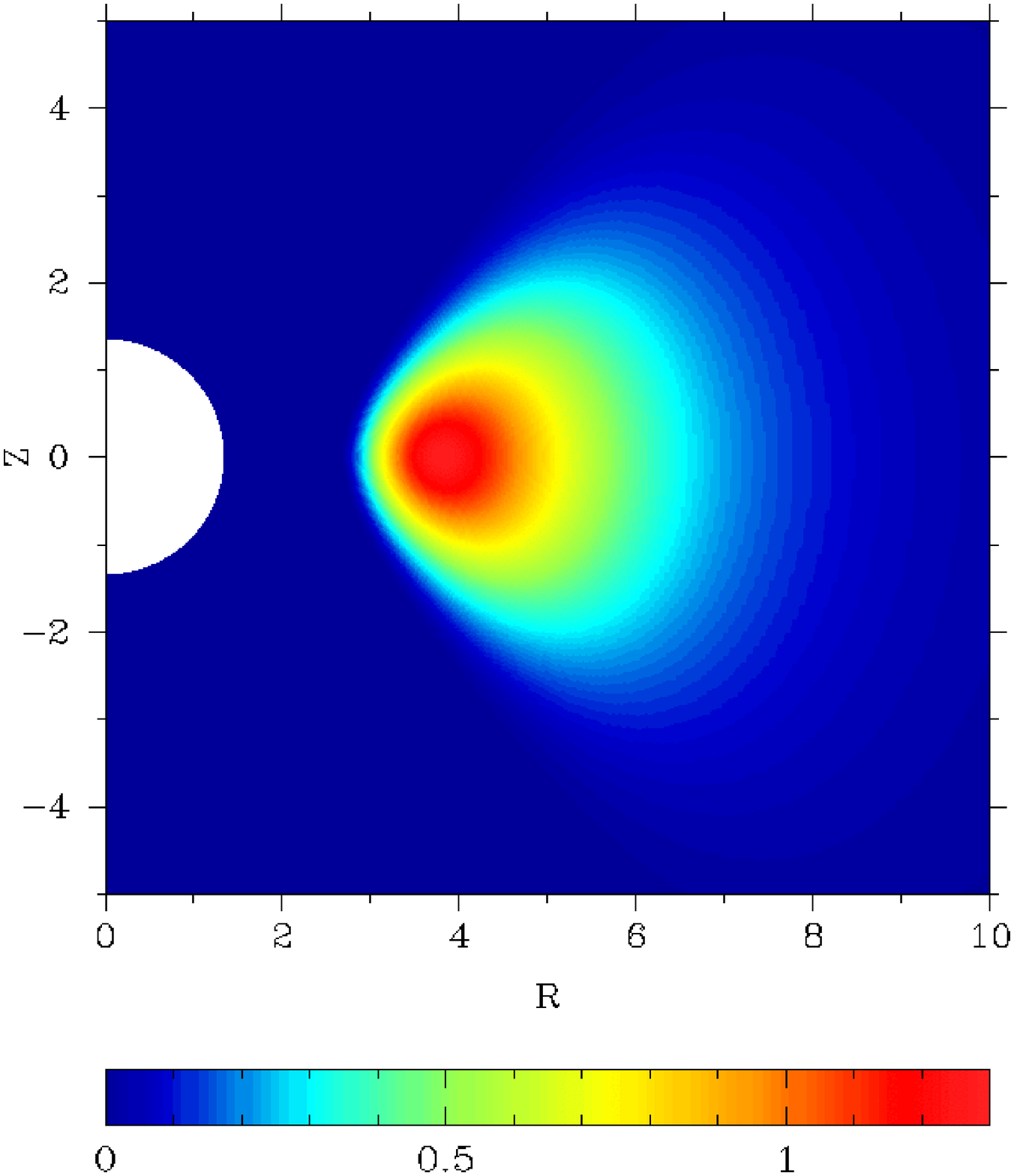}}
\caption{ Exact solution for model A. The horizontal coordinate 
is $R=r \sin\theta$ and the vertical coordinate is $Z=r\cos\theta$ 
{\it Left panel:} $\beta$, 
{\it Middle panel:} $\Omega$, 
{\it Right panel:} $\rho$.
}
\label{model-a}
\end{figure*}

\begin{figure*}
\fbox{\includegraphics[width=55mm]{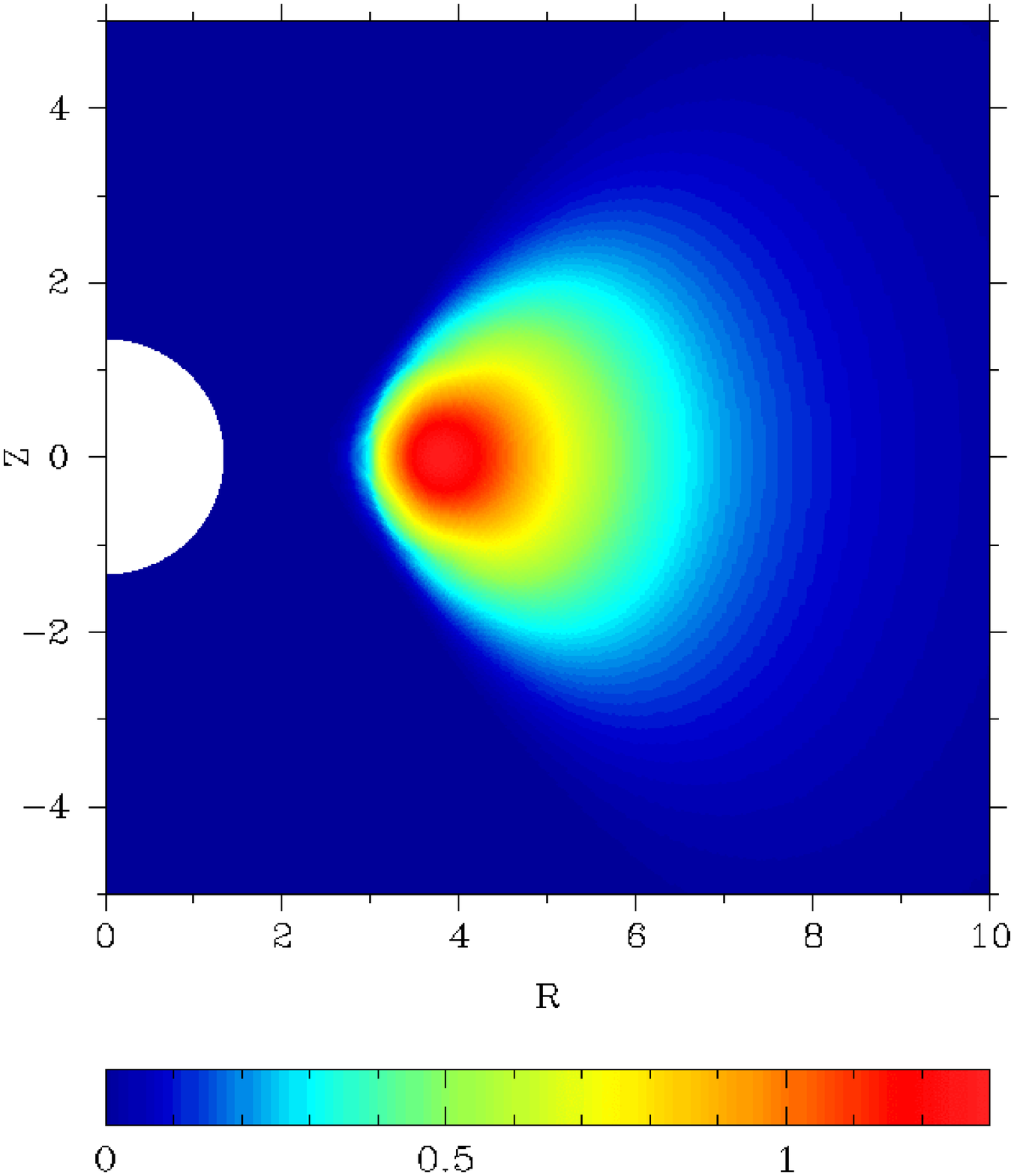}}
\fbox{\includegraphics[width=55mm]{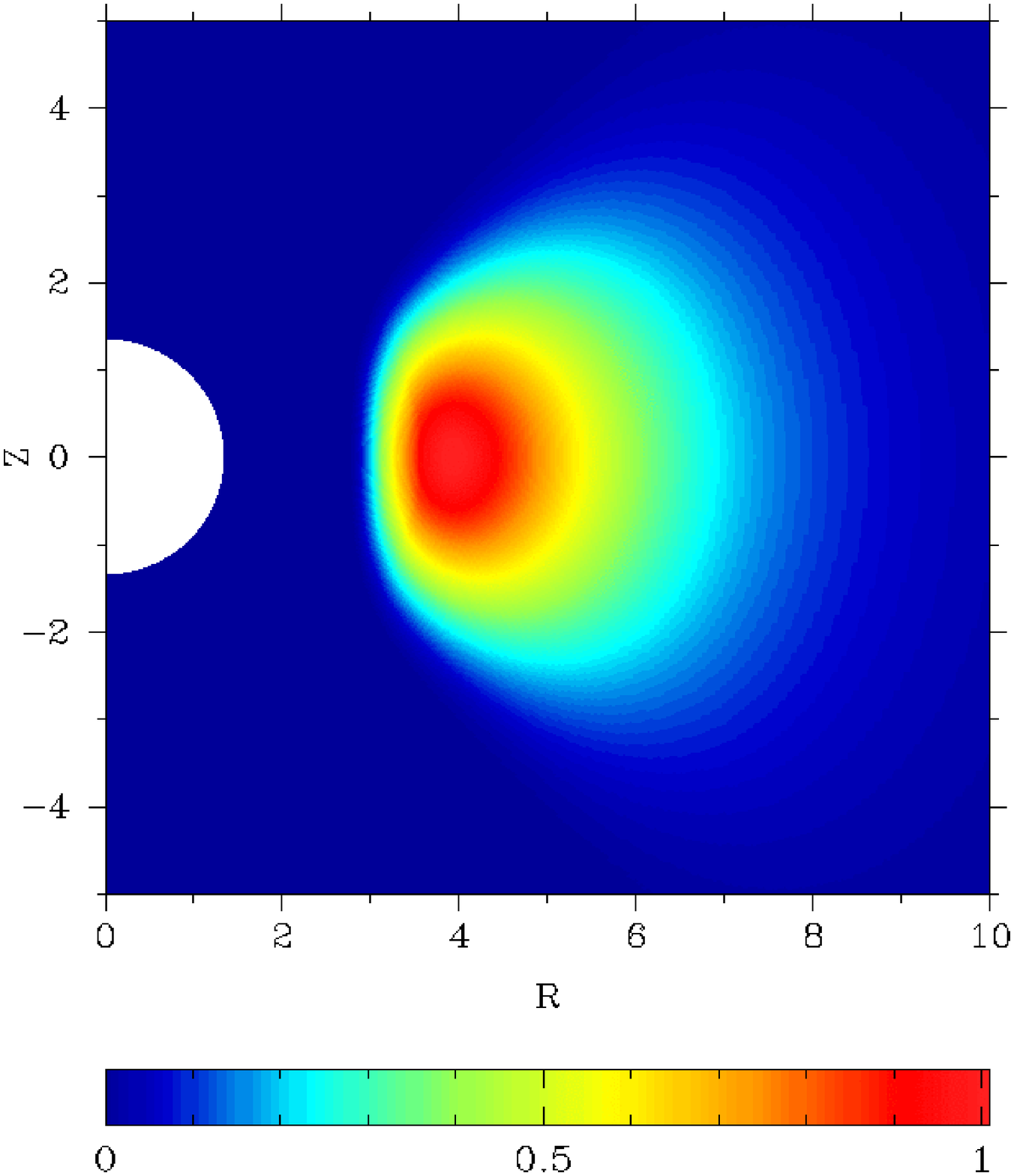}}
\fbox{\includegraphics[width=55mm]{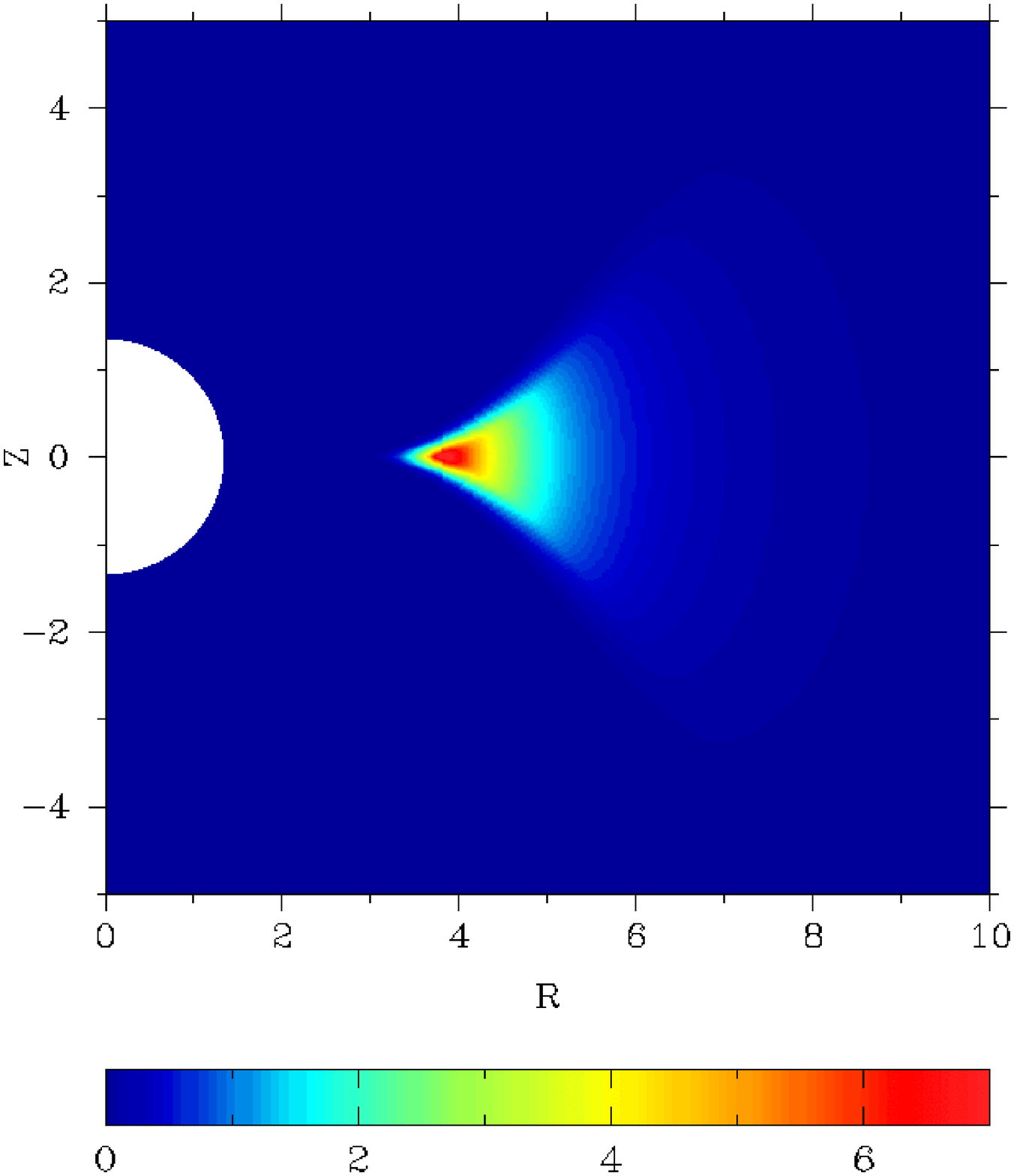}}
\caption{The density distributions for the model A obtained with 
our actual numerical code and its corrupted versions. In all these
plots the linear scaling is used.   
{\it Left panel:} the solution obtained with the uncorrupted code at 
$t=200$.
{\it Middle panel:}  the solution obtained with the corrupted version of 
the code at $t=14$. Here we retained only the contribution of magnetic 
pressure in the Maxwell stress tensor. 
{\it Right panel:} the solution obtained with the corrupted version of
the code at $t=14$. Here all the electromagnetic contributions 
to the stress-energy-momentum tensor were omitted.
}
\label{referee}
\end{figure*}
                                                                                        
\subsection{Simulations}

In order to illustrate the usefulness of this solution for testing 
GRMHD computer codes we constructed two equilibrium models and 
used them to setup the initial solution for 2D axisymmetric simulations 
using the code described in Komissarov \shortcite{Kom}. In both models the 
black hole has specific angular momentum $a=0.9$ which gives $r_{mb}=1.73$, 
$r_{ms}=2.32$, $l_{mb}=2.63$, $l_{ms}=2.49$. In both cases, we used   
the same value for the barotropic powers $\kappa,\eta=4/3$ and 
the polytropic equation of state with the same ratio of specific 
heats, $\gamma=4/3$.  The other parameters are given in Table 1. 
Notice, that $l_0>l_{mb}$ in model A, whereas in model B one has 
$l_{ms}<l_0<l_{mb}$ and this allows accretion through the torus 
cusp \cite{Abr78,Koz}. 

Initially, the space outside of the tori is filled with a rarefied non-magnetic 
plasma accreting into the black hole. Its density and pressure are  
$$
  \rho = 10^{-3} \rho_c \exp(-3r/r_c),\quad  
   p = K\rho^{\kappa}. 
$$  
Its velocity in the frame of local fiducial observer, FIDO, \cite{TM} is
radial and has the magnitude 
$$
   v=\beta^{\hat{r}} (1-(r_g/r)^4) 
$$  
where $\beta^{\hat{r}}$ is the radial component of velocity of the spacial grid 
relative to  FIDO. This introduces inflow through the horizon with the local 
velocity of FIDOs and at the same time allows only small poloidal velocity jump 
at the torus surface. The latter property allows to avoid  strong rarefactions
that may originate at the torus surface right at the start of simulations.  
The initial distributions of $\rho$, $\Omega$, and $\beta$ for the model A, 
that is a more extreme case, are shown in figure \ref{model-a}. Model B 
appears very similar.     

In these simulations we utilized the Kerr-Schild coordinates $(r,\theta)$. 
The computational domain is $[ 1.35,53.3]\times [0,\pi]$ with 320 cells in 
each direction.  At lower resolution the effects of numerical diffusion become 
rather noticeable.  The grid is uniform in the $\theta$-direction and the cell 
size in the radial direction is such that in the equatorial plane 
$g_{\theta\theta}\Delta\theta^2 = g_{rr}\Delta r^2$.
This ensures that computational cells have approximately equal lengths in 
both directions and throughout the whole grid. 

\begin{table}        
   \begin{tabular}{|c|l|l|l|l|l|l|l|}
   \hline
    Model & $l_0$ & $r_{cusp}$ & $r_c$ & $W_{cusp}$ & $W_{c}$ & $W_{in}$ & $\beta_c$ \\
   \hline
     A & 2.8 & 1.58 & 4.62 &  0.702 & -0.103 & -0.030 & 0.1   \\  
   \hline
     B & 2.6 & 1.78 & 3.40 & -0.053 & -0.136 & -0.053 & 1.0 \\ 
   \hline
   \end{tabular}
\caption{Parameters of the models used for test simulations}
\end{table}

The rotational period of the disc centre is rather long $\tau_r=68$ in model 
A and $\tau_r=45$ in  model B. However, the dynamical timescale, $\tau_d$, for the 
disc centre is significantly shorter. Indeed, in both of these models the fast 
magnetosonic speed in the disc centre is $a_f \simeq 0.18$, whereas the 
length scale of pressure distribution at this location is $L_c \simeq 1.0$. 
This gives the dynamical timescale $\tau_d \simeq 6 $. Since it is the dynamical
timescale that determines how quickly the system reacts to perturbations of 
its equilibrium state,  it is quite sufficient to carry out simulations for 
only $t= \mbox{few } \tau_d$ in order to test the ability of our numerical code 
to correctly reproduce these equilibrium solutions. In order to demonstrate 
this we have carried out test simulations for model A not only with our ``proper code'' 
but also with its two corrupted versions. In the first version, it is only the 
electromagnetic
pressure, $(B^2+E^2)/2$, that was taken into account in the calculations of 
the Maxwell stress tensor. In the 
second and more drastic version we omitted all the contributions of the 
electromagnetic field to the stress-energy-momentum tensor. The results are 
presented in figure \ref{referee}. One can see that already at 
$t \simeq 2 \tau_d$ the solutions 
obtained with the corrupted versions are significantly different from the 
initial equilibrium solution. With the second corrupted version the disc simply 
collapses towards the equator due to the lack of magnetic support against gravity. 

\begin{figure*}
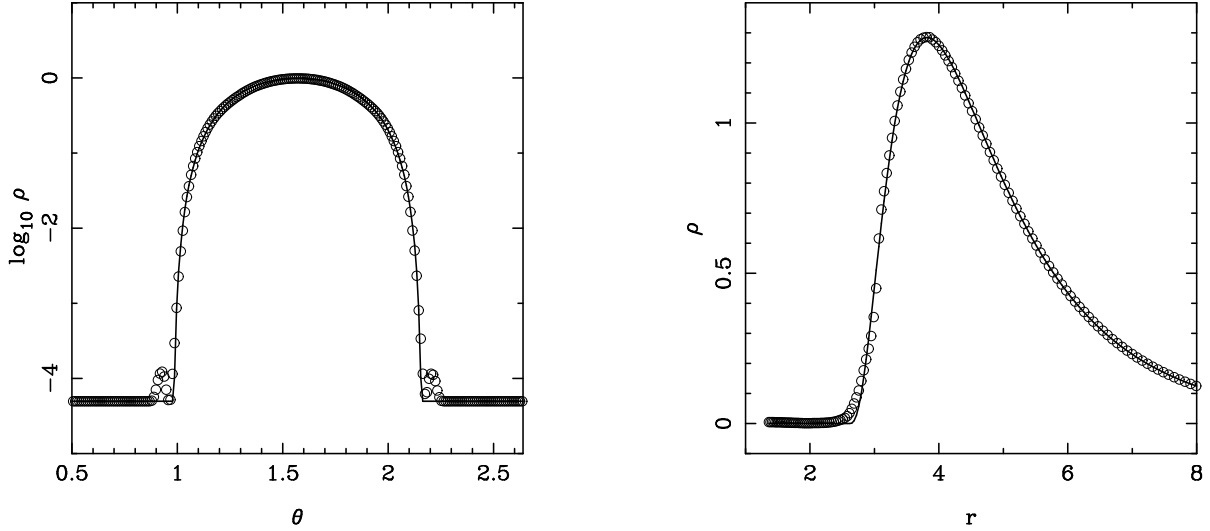

\includegraphics[width=70mm,angle=-90]{figures/t-1.ps}
\hskip 2cm 
\includegraphics[width=70mm,angle=-90]{figures/t-2.ps}
\caption{Density distribution at $t=60$ for model A. The solid lines 
show the exact equilibrium solution.  {\it Left panel:} $\log_{10}\rho$ 
against polar angle $\theta$ at $r=r_c$, the disc centre location. 
{\it Right panel:} $\rho$ against $r$ in the equatorial plane.
}
\label{comp-A}
\end{figure*}

\begin{figure*}
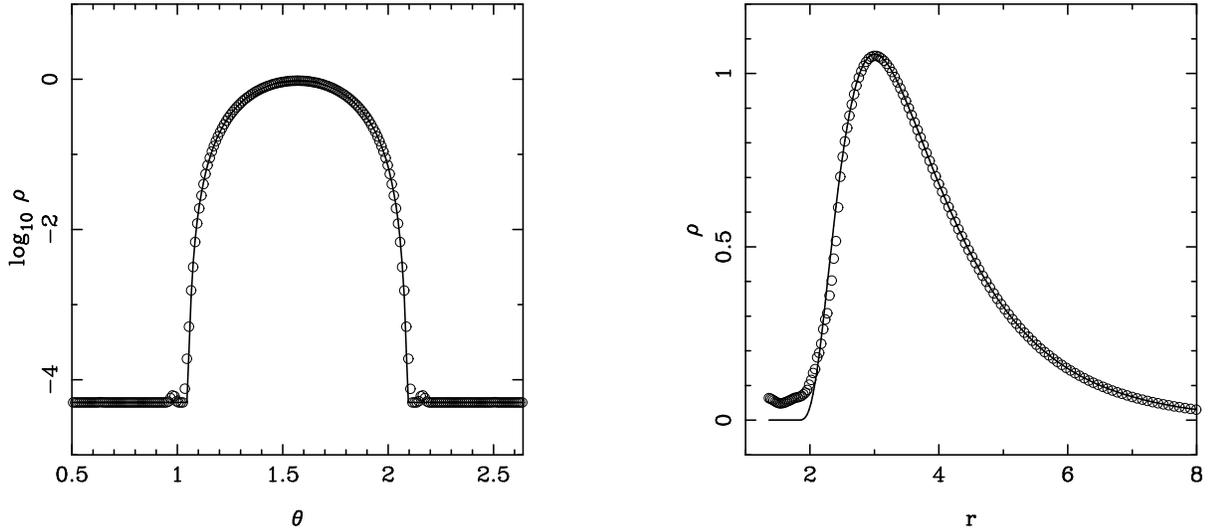

\includegraphics[width=70mm,angle=-90]{figures/t1-1.ps}
\hskip 2cm 
\includegraphics[width=70mm,angle=-90]{figures/t1-2.ps}
\caption{Density distribution at $t=60$ for model B. The solid lines
show the exact equilibrium solution.  {\it Left panel:} $\log_{10}\rho$
against the polar angle $\theta$ at $r=r_c$, the disc centre location.
{\it Right panel:} $\rho$ against $r$ in the equatorial plane.
}
\label{comp-B}
\end{figure*}

Figure \ref{referee} also presents the proper 
numerical solution at t=200, which is $30\div 40$ times larger than $\tau_d$. 
``Naked eye'' inspection shows that this solution is very similar to the initial 
equilibrium one that is presented in figure \ref{model-a}.
This is confirmed by the 1D density plots showing in details the 
distributions along and across the symmetry plane: figure \ref{comp-A} 
for model A and figure \ref{comp-B} for model B. Other images also reveal 
what appears to be surface waves propagating away from the black hole and becoming 
noticeable in the more remote parts of the tori.

\section{Summary} 

In this paper we have generalized the relativistic theory of thick accretion 
discs around Kerr black holes \cite{FiM,Abr78,Koz} by including dynamically 
strong toroidal (azimuthal) magnetic field. As expected, this inclusion of 
magnetic field leads to a whole new class of equilibrium solutions
that differ in strength and spacial distribution of the field. 
In particular, we have described the way of constructing barotropic tori 
with constant angular momentum -- under such conditions the differential equations 
of magnetostatics are easily integrated and reduce to simple algebraic equations. 

By now it has become clear that magnetic field plays many important roles
in the dynamics of astrophysical black hole-accretion disc systems. However, 
the structure of this field may be more complex than just azimuthal loops 
and as the result the analytic solutions constructed in this paper may turned out 
not to be particularly suitable for modelling real astrophysical objects. 
Even so they will be certainly helpful in testing computer codes for general 
relativistic MHD that are becoming invaluable tools in the astrophysics of 
black holes/accretion discs.

\section*{Acknowledgements}
 This research was funded by PPARC under the rolling grant
``Theoretical Astrophysics in Leeds''

\appendix
\section{ }

Here we derive equations (\ref{eq-1}) and (\ref{eq-4}) of the 
main paper.

\subsection{Equation (\ref{eq-1})}

The contraction of the energy-momentum equation,  
$$
 \nabla_\alpha T^{\alpha \beta} = 0,
$$
with the projection tensor 
$$
h^{\alpha}_{\ \beta} = \delta^{\alpha}_{\ \beta} +u^\alpha u_\beta   
$$ 
leads to 

\begin{equation}
   \begin{array}{rcl}
  h^{\alpha}_{\ i} \nabla_\gamma \left( (w+b^2) u^\gamma u_\alpha \right) &+& \\ 
  h^{\alpha}_{\ i} \nabla_\gamma \left( (p+b^2/2)\delta^\gamma_{\ \alpha}\right) &+& \\ 
  h^{\alpha}_{\ i} \nabla_\gamma \left( b^\gamma b_\alpha\right)  &=& 0, 
\end{array}
\label{eq-a1}
\end{equation}

\noindent
where $i=r,\theta$. Applying the product rule and using that 
$h^{\alpha}_{\ \beta} u_\alpha=0$ and 
$u^\alpha\nabla_\gamma u_\alpha=0$ one can write 

$$
 h^{\alpha}_{\ i}\nabla_\gamma\left((w+b^2) u^\gamma u_\alpha\right) = 
    (w+b^2)\left[ \nabla_\gamma(u^\gamma u_i)  - 
                  u_i \nabla_\gamma u^\gamma   \right].  
$$

\noindent
The second term in the square brackets vanishes because of the symmetry 
conditions (\ref{Kerr-m}-\ref{azim-b}). Applying the well known result for the 
divergence of a second rank symmetric tensor on obtains   

$$
   \nabla_\gamma(u^\gamma u_i) = \frac{1}{\sqrt{-g}}(\sqrt{-g}u^\gamma u_i)_{,\gamma}
    -\frac{1}{2}g_{\mu\nu,i}u^\mu u^\nu 
$$

\noindent
The first term in this equation vanishes because of the same symmetry conditions 
whereas 
$$
  g_{\mu\nu,i}u^\mu u^\nu = -2u_\nu u^\nu_{\ ,i}.
$$ 

\noindent
Thus, one has 
                     
\begin{equation}
    h^{\alpha}_{\ i}\nabla_\gamma\left((w+b^2) u^\gamma u_\alpha\right) = 
    (w+b^2)u_\nu u^\nu_{\ ,i}.
\label{eq-a2}
\end{equation}

\noindent
Now we deal with the second term in (\ref{eq-a1}).  
\begin{eqnarray}
\nonumber
 h^{\alpha}_{\ i} \nabla_\gamma \left( (p+b^2/2)\delta^\gamma_{\ \alpha}\right) &=& 
 h^{\gamma}_{\ i} \nabla_\gamma (p+b^2/2)  \\
\nonumber
 &=&  (p+b^2/2)_{,i} + u_i u^\gamma (p+b^2/2)_{,\gamma}.
\end{eqnarray}
The second term on the right-hand side of this equation vanishes because of the 
symmetries and hence we have

\begin{equation}
 h^{\alpha}_{\ i} \nabla_\gamma \left( (p+b^2/2)\delta^\gamma_{\ \alpha}\right) = 
(p+b^2/2)_{,i}.
\label{eq-a3}
\end{equation}

\noindent
As to the third term in  (\ref{eq-a1}),  
\begin{eqnarray}
\nonumber
h^{\alpha}_{\ i} \nabla_\gamma \left( b^\gamma b_\alpha\right) &=& 
    h^{\alpha}_{\ i}\left( \frac{1}{\sqrt{-g}}(\sqrt{-g}b^\gamma b_\alpha)_{,\gamma}  
    -\frac{1}{2}g_{\mu\nu,\alpha} b^\mu b^\nu \right)  \\
\nonumber
   &=& -\frac{1}{2} h^{\alpha}_{\ i}g_{\mu\nu,\alpha} b^\mu b^\nu  \\
\nonumber
   &=& -\frac{1}{2}\left( g_{\mu\nu,i} b^\mu b^\nu   
                         +u_i u^\alpha g_{\mu\nu,\alpha} b^\mu b^\nu   \right)  \\
\nonumber
  &=& -\frac{1}{2}g_{\mu\nu,i} b^\mu b^\nu  \\
\label{eq-a4}
  &=&  -\frac{1}{2}( b^2_{\ ,i} -2 b_\nu b^\nu_{\ ,i}).  
\end{eqnarray}
During this reduction we twice used the symmetry conditions. 
Substituting of results (\ref{eq-a2}-\ref{eq-a4}) into eq.(\ref{eq-a1}) 
leads to 
\begin{equation}
    (w+b^2)u_\nu u^\nu_{\ ,i} +
    (p+b^2)_{,i} - b_\nu b^\nu_{\ ,i} =0, 
\label{eq-a5}
\end{equation}
which is equation (\ref{eq-1}) of the main paper. 

\subsection{Equation (\ref{eq-4})}

From the definitions of the angular velocity, $\Omega$, and the 
angular momentum, $l$, and the symmetries of the problem it immediately 
follows that 

\begin{equation}
   l= -\frac{g_{t\phi} + g_{\phi\phi}\Omega}
            {g_{tt} + g_{t\phi}\Omega  },
\label{eq-a6} 
\end{equation}

\begin{equation}
   \Omega= -\frac{g_{t\phi} + g_{tt} l}
            {g_{\phi\phi} + g_{t\phi} l },
\label{eq-a7}
\end{equation}
where we assumed the Kerr metric in either Kerr-Schild or 
Boyer-Lindquist coordinates. From the definition of 4-velocity it 
follows that 
$$
   g_{\mu\nu}u^\mu u^\nu=-1. 
$$
In the case under consideration this leads to 

\begin{equation}
    (u^t)^2 = -(g_{tt}+2g_{t\phi}\Omega + g_{\phi\phi}\Omega^2)^{-1}, 
\label{eq-a8}
\end{equation}
and 
\begin{equation}
    u^t u_t = -\frac{1}{1-l\Omega}
\label{eq-a8a}
\end{equation}
From (\ref{eq-a7}-\ref{eq-a8a}) one finds 
                                                                                              
\begin{equation}
    (u_t)^2 = \frac{\cal L}{\cal A},
\label{eq-a9}
\end{equation}
where 

\begin{equation}
    {\cal L}  = g_{t\phi}g_{t\phi} -g_{tt}g_{\phi\phi},
\label{eq-a10}
\end{equation}
and 
\begin{equation}
    {\cal A}  = g_{\phi\phi}+2lg_{t\phi}+ l^2 g_{tt}.
\label{eq-a10a}
\end{equation}

\noindent
From the constraint equation (\ref{constraint}) and the conditions 
(\ref{azim-u},\ref{azim-b}) one finds                               
                     
\begin{equation}
   b^t=l b^\phi,
\label{eq-a11}
\end{equation}

\begin{equation}
   b_t=-\Omega b_\phi,
\label{eq-a12}
\end{equation}

\noindent
These allow to write
\begin{equation}
   b^2=b_\phi b^\phi (1-l\Omega),
\label{eq-a13}
\end{equation}
and
\begin{equation}
   b^2=(b^\phi)^2 {\cal A}.
\label{eq-a13a}
\end{equation}
\noindent
Now we reduce the terms $u_\nu u^\nu_{\ ,i}$ and 
$b_\nu b^\nu_{\ ,i}$ in  equation (\ref{eq-a5}). Using the definitions of
$\Omega$ and $l$ one derives  
\begin{eqnarray}
\nonumber
    u_\nu u^\nu_{\ ,i} &=& - u^\nu u_{\nu,i} \\
\nonumber
    &=& -u^t u_{t,i} -u^\phi u_{\phi,i} \\
\nonumber
    &=& -u^t \left(u_{t,i}-\Omega (lu_t)_{,i}\right)  \\
\nonumber
    &=& -u^t \left(u_{t,i}(1-l\Omega) -\Omega u_t l_{,i}\right).
\end{eqnarray}
The substitution of $u^t$ from (\ref{eq-a8a}) into the last equation 
gives us 
 
\begin{equation}
    u_\nu u^\nu_{\ ,i} = (\ln{|u_t|})_{,i} - \frac{\Omega}{1-l\Omega} l_{,i}
\label{eq-a14}
\end{equation}

\noindent
Using (\ref{eq-a11}) and (\ref{eq-a12}) one writes
\begin{eqnarray}
\nonumber
  b_\nu b^\nu_{\ ,i} &=& b_tb^{t}_{\ ,i}+ b_\phi b^{\phi}_{\ ,i} \\
\nonumber
        &=& -\Omega b_\phi (lb^\phi)_{,i} +b_\phi b^{\phi}_{\ ,i} \\
\nonumber
        &=& b_\phi b^\phi_{\ ,i} (1-l\Omega) - \Omega b_\phi b^\phi l_{,i}.
\end{eqnarray}
The substitution of $b_\phi$ from (\ref{eq-a13}) into this equation gives 
                                                                                              
\begin{equation}
    b_\nu b^\nu_{\ ,i} = b^2 (\ln |b^\phi|)_{,i} -\frac{\Omega b^2}{1-l\Omega} l_{,i}.
\label{eq-a15}
\end{equation}
                                                                                              
\noindent
Substituting (\ref{eq-a14},\ref{eq-a15}) into (\ref{eq-a5}) one finds 

\begin{eqnarray}
\nonumber
    w \left(  (\ln{|u_t|})_{,i} - \frac{\Omega}{1-l\Omega} l_{,i}  \right)  
    + p_{,i} \\ 
    + b^2 \left( \ln{\left|\frac{u_t}{b^\phi}\right|} \right)_{,i} 
    + b^2_{,i} &=& 0 
\label{eq-a16}
\end{eqnarray}
                                                                                              
\noindent
Using (\ref{eq-a9}) and (\ref{eq-a13a}) one obtains

$$
b^2 \left( \ln{\left|\frac{u_t}{b^\phi}\right|} \right)_{,i} + b^2_{,i}  
  = \frac{({\cal L} b^2)_{,i}}{2 {\cal L}}.  
$$
Thus, (\ref{eq-a16}) can be written as 
\begin{equation}
    (\ln{|u_t|})_{,i} - \frac{\Omega}{1-l\Omega} l_{,i}  
    + \frac{p_{,i}}{w} + \frac{({\cal L} b^2)_{,i}}{2 {\cal L} w} =0,
\label{eq-a17}
\end{equation}
which is equation (\ref{eq-4}) of the main paper.


\end{document}